\documentclass[twocolumn,prl,showpacs,preprintnumbers,amsmath,amssymb,a4paper,fleqn]{revtex4-1}

\usepackage{dcolumn}
\usepackage{epsfig}
\usepackage{amssymb,amsmath,bm,graphicx,multirow,mathtools}
\usepackage{color}

\def\be{\begin{equation}}
\def\ee{\end{equation}}
\def\bea{\begin{eqnarray}}
\def\eea{\end{eqnarray}}
\def\beeg{\begin{align}}
\def\eeg{\end{align}}

\def\m{\mu}
\def\n{\nu}

\def\p{\partial}
\def\a{\alpha}
\def\b{\beta}
\def\t{\theta}
\def\g{\gamma}

\def\l{{\lambda}}

\def\11{1\hspace{-0.6em}1}
\def\22{1\hspace{-0.48em}1}

\def\yh{\hat{y}}

\def\i{\imath}

\begin{document}

\title{Non-commutativity on another Minkowski space-time: Vierbein formalism and Higgs approach
%: Vierbein formalism and Inverse mapping from $T_pM$
}
\author{Abolfazl\ Jafari\footnote{jafari-ab@sci.sku.ac.ir}}
\affiliation{Department of Physics, Faculty of Science,
Shahrekord University, P. O. Box 115, Shahrekord, Iran}
\date{\today }

\begin{abstract}

\textbf{Abstract:} 
We extend the non-commutative coordinates relationship into other than the Minkowski space-time. We clarify the non-commutativity dependency to the geometrical structure.  
As well as, we find an inverse map between Riemann's normal and global coordinates.
Furthermore, we show that behavior of the corresponding coordinates non-commutativity like as a tensor.
All results summarized for the Schwarzschild metric.  
And they summarized for the space-time in the presence of weak gravitational waves.

\end{abstract}


\maketitle


\section{Introduction}

Space-time at local is flat and the asymptotic conception of a more general space-time.
It also has a correlation characteristic.
%It's also about the concept of correlation.
When a pseudo-Riemann's manifold gets the notion of correlation, then a system of measurements is available.
%For every event $'p'$ in a space-time, each enough nearby neighborhood of $'p'$ can always be spanned by local coordinates if there is a unique geodesic %joining between any point of the neighborhood and $'P'$.
One can always span each neighborhood of an event $'p'$ in space-time by local coordinates 
if it is possible to exist a unique geodesic joining between any points of the neighborhood and $'P'$. 
Therefore, any Riemannian manifold equipped with normal coordinates \cite{lab01, lab02}.
%Each local framework is associative with an asymptotic space.
Minkowski space-time has an asymptotic conception for any solutions to Einstein's equation which is descriptive of a free-falling frame in this notion.
Minkowski space-time always is available if the measurements are in the inertial frame.
%55The tools such as tetrad fields make an ability asymptotic state from a base space-time.
%Tetrad fields are responsible for this mapping.
Vierbein field is a tool for generalizing a theorem from the tangential space to its main one.
%
%The mapping $\mathit{T}_p\mathit{M}\rightarrow\mathit{M}$ is in the first order of the affine connections.  
%There is a reversible manner.
%Thus 
%We construct vierbein fields as a transformation matrix between global and local coordinates or as a set of $ "d" $-vectors in $\mathit {T}_p\mathit{M} $.
In general relativity, a vierbein field is a set of four orthonormal vector fields. 
%The timelike unit vector field is often denoted by ${\displaystyle {\vec {e}}_{0}} \vec{e}_0$ and the three spacelike unit vector fields by ${\displaystyle {\vec {e}}_{1},{\vec {e}}_{2},\,{\vec {e}}_{3}} \vec{e}_1, \vec{e}_2, \, \vec{e}_3$. 
Express of all tensorial quantities on a manifold is by using the frame field and its dual co-frame field.
The set of the vierbein field contains $"d"$-vector fields. Their numbers are proportional to the dimension of space-time.
One of them is time-like while rests are space-like. 
They are on a Lorentz manifold and interpreted as a model of space-time.
It is important to recognize that frame fields are geometric quantities. They made based on the metric tensor of a general space-time.
Vierbein fields make sense independent of a choice of a coordinate chart. They do the notions of orthogonality and length.
%Thus, just like vector fields and other geometric quantities, frame fields can be represented in various coordinate charts. 
%Computations will always yield the same result between the components of tensorial quantities, with respect to a given frame,
%will always yield the same result, whichever coordinate chart is used to represent the frame.
%when the coordinate chart is used to represent the frame.
%{\color{blue}
%These fields are required to extract curved space time version of the Minkowskian physical quantities.
Therefore, vierbein formalism is an official approach. By these, Minkowski version of some physical quantities generalizes into a general space-time.
%They are made based on the metric tensor of general space-time.
%}
%By exploiting vierbein fields; $a^\a_{\ \m}$, we can find the physical quantities on the pseudo Riemann manifolds based on the values of them on relevant %tangent space.
Vierbein field $a^\a_{\ \m}$ has two kinds of indices. 
$\m$ indicates the coordinates of a general space-time and $\a$ is only a counter.
Changes in $\m$ occur by the metric tensor while the number of vierbein fields enumerated by $\a$.
%indices $\a$, $\b$ act merely as labels.
%Under transformations of the coordinates, indices $\m$, $\n$, are regarded as tensor
%indices, while indices $\a$, $\b$ act merely as labels. 
The formalism is invariance under the Lorentz transformation $\a$ as well as $\m$.
Vierbein counters are rising by $\eta_{\a\b}$, while the space-time indices with the metric $g_{\m\n}$.
%For vierbein fields, $a^\a_{\ \m}$, $\a$ as a upper index is a label and takes the values from $\{1,2,\cdots,d\}$.
%Also, the bottom index $\m$, is a Lorentzian index and runs from 0 to d.
Lorentz indices run from zero to $"\mathit{d}"$, dimension of space-time \cite{lab01, lab03, lab04}.
%Of course, this approach is efficient on the tangent spaces.
%Developing via vierbein fields reach to an approximation theory where the values of Riemann's curvature tensors are perturbation parameter.
%Perturbation parameter is calculated in term of values of Riemann's tensors at any event as a point belonging to the hypersurface of space-time. 
%Frame fields were already used to make coordinates dependent on $\gamma^\m$-matrices.
%Therefore, we have to provide more set of vierbein fields. 
Vierbein field is as a square root of the metric tensor \cite{lab01, lab02, lab03, lab04, lab05}
\begin{align}\label{vierbein-fields}
g_{\m\n}(x)=a^{\a}_{\ \m}(y)\eta_{\a\b}a^{\b}_{\ \n}(y).
\end{align}
%in which under coordinate transformation indices, $\m$ and $\n$ are regarded as tensor indices while indices of $\a$ and $\b$ act merely as %labels.
Clearly, in the whole of a manifold, the vierbein field given by $a^\a_{\ \m}(y)=\delta^\a_{\ \m}$ when gravity is absent.
There is another way to find the non-commutativity dependency to the geometrical characteristics.
This way named Higgs approach.   
For a small neighborhood of $'P'$ in $\mathit{M}$, global and local coordinates related to each other. 
Also, the local metric of this region achieves by employing the Higgs approach. 

We deal with the non-commutative coordinates when we attempt to make of change in physics for short distances \cite{lab06}.
%We admit of new conception from coordinates by reporting theoretical qunsquance of length   
%Non-commutativity relationship occurs between the coordinates of the same frame.

We can introduce the non-commutative framework with a replacement of Hermitian operators $\yh^\m$ instead of local coordinates $y^\m$. 
So that obeys the following relation
\begin{align}\label{gnc}
[\yh^\m,\yh^\n]=\i\tilde{\t}^{\m\n}(\yh),
\end{align} 
which for Minkowski space-time, Eq.(\ref{gnc}) gets the following form
\begin{align}\label{mnc}
[\yh^\m,\yh^\n]=\i\t^{\m\n}
\end{align} 
$\t^{\m\n}$ is a constant and real parameter \cite{lab06, lab07, lab08, lab09, lab10, lab11, lab12}.
In the limit of $\mathit{T}_p\mathit{M}$, the results of the non-commutative version of physics depend on the choice of an event in the space-time. 
For, we determine $\tilde{\t}(\yh)$ up to the first order of the Riemann curvature tensor.
%$\tilde{\t}(\yh)$ should be known although approximately with the Riemann curvature tensor an approximation parameter.
Since many properties of tangential spaces induced from the base manifolds direct. 
So, the non-commutative relationship on $\mathit{T}_p\mathit{M}$ comes from $\mathit{M}$.
Obviously that why we are studying for the main non-commutativity of coordinates on the $\mathit{M}$.

The purpose of this paper is to recognize the dependence of $\tilde{\t}$ on coordinates.
We also find another mapping other than vierbein formalism. 
So that, it helps us to get back the Minkowskian coordinates non-commutativity into the original manifold.
Then, we compare the results of the two proposed methods.

\section{Presentation of the theory}
We make two sets of vierbein fields based on the given metrics, the metric of weak gravitational waves and Schwarzschild universe.
We assume an event of space-time $"P"$ (with the coordinate $'Y_p'$). Suppose that its relevant geometry quantities are slowly varying functions.
Local and global coordinates ($y$ and $x$) span the pseudo-Riemann manifolds in the partial and whole states. 
%%%%%The special relativity can be derived based on local coordinates. 
%are made based on the framework built on the manifolds.
For a point $'Y_p'$ of space-time, there is a fundamental relationship linking them to one another
\begin{align}\label{eq1}
x^\m=\cr
Y^\m_p+y^\m+A^\m_{\a\b}(Y_p)y^\a y^\b
+B^\m_{\a\b\g}(Y_p)y^\a y^\b y^\g
+\cdots
%\cr
%-\Gamma^\m_{\a\b}(Y_p)y^\a y^\b+\cdots.
\end{align}
$A^\m_{\a\b}$ and $B^\m_{\a\b\g}$ are the Christopher coefficients and their derivatives. 
The amount of them calculated at the event point.
The set of coordinates $'Y_p'$ indicates a global framework. Also the validity of Eq.(\ref{eq1}) restricted to a small neighborhood of $\mathit{T}_p\mathit{M}$.
Obviously that, the variation of the general coordinate $'Y_p'$ is small compared to the local coordinate $'y'$.
% in which special relativity can be derived based on local coordinates \cite{lab01, lab02}.
Nevertheless, metric expansion becomes \cite{lab01, lab03}
\begin{align}\label{eq2}
g_{\a\b}(y)=g_{\a\b}(Y_p)-\frac{1}{3}R_{\a\m\b\n}(Y_p)y^\m y^\lambda
\cr
-\frac{1}{3!}R_{\a\gamma\b\n;\m}(Y_p)y^\n y^\m y^\gamma+\cdots,
\end{align}
of course, $g_{\a\b}(Y_p)\approx\eta_{\a\b}$, and
\begin{align}\label{eq3}
\Gamma^\a_{\b\gamma}(y)=\Gamma^\a_{\b\gamma}(Y_p)+\p_\m\Gamma^\a_{\b\gamma}(Y_p)y^\m+\cdots
\end{align}
Hence, at the origin of a framework, Riemann's curvature tensors will be
\begin{align}\label{eq4}
R^\a_{\b\m\n}(Y_p)=\Big{(}\p_\n\Gamma^\a_{\b\m}-\p_\m\Gamma^\a_{\b\n}\Big{)}\mid_{\mathrm{calculated\ at\ Y_p}}
\end{align}

\textbf{Static case of $\mathit{T}_p\mathit{M}$}- Schwarzschild universe is a static solution to Einstein's equation
%Many static solutions to the general relativity equation are available such as Schwarzschild solution 
\cite{lab04, lab13}.
We assume that $\mathit{T}_p\mathit{M}$ is a falling space along a geodesic of Schwarzschild space-time.
Hyper-surface of space-time with constant time is perpendicular to the geodesic.
So, its affine connections will be time-independent \cite{lab01,lab03}.
Based on Eq.(\ref{eq2}), we provide a set of vierbein fields which specialized for Schwarzschild universe. 
%Metric tensor is given by Eq.(\ref{eq2}) and due to the fact that we provide a set of vierbein fields.
In matrix representation, vierbein fields found to be: \cite{lab14, lab15}
\begin{widetext}
\begin{align}\label{amiu2}
a^\a_{\ \m}=\left(\begin{array}{cccc}
                 1-\frac{1}{2}R^0_{\ l0m}y^ly^m & -\frac{1}{6}R^0_{\ l1m}y^ly^m & -\frac{1}{6}R^0_{\ l2m}y^lx^m & -\frac{1}{6}R^0_{\ l3m}y^ly^m  \\
                 -\frac{1}{2}R^1_{\ l0m}y^ly^m & 1-\frac{1}{6}R^1_{\ l1m}y^ly^m & -\frac{1}{6}R^1_{\ l2m}y^ly^m & -\frac{1}{6}R^1_{\ l3m}y^ly^m  \\
                 -\frac{1}{2}R^2_{\ l0m}y^ly^m & -\frac{1}{6}R^2_{\ l1m}y^ly^m &  1-\frac{1}{6}R^2_{\ l2m}y^ly^m & -\frac{1}{6}R^2_{\ l3m}y^ly^m  \\
                 -\frac{1}{2}R^3_{\ l0m}y^ly^m & -\frac{1}{6}R^3_{\ l1m}y^ly^m &  -\frac{1}{6}R^3_{\ l2m}y^ly^m & 1-\frac{1}{6}R^3_{\ l3m}y^ly^m
               \end{array}
\right).
\end{align}
\end{widetext} 
which simplify to the following expression
\bea\label{amiu3}
a^\a_{\ \m}=\delta^\a_\m-\frac{\xi^\a_\m}{6}R^\a_{\ l\m k}y^ly^k,
\eea
where $\xi^\a_i=1$ and $\xi^\a_0=3$.
To get the coordinates non-commutative relationship on the $\mathit{M}$ from the version of $\mathit{T}_p\mathit{M}$, 
%we keep terms only to the first order of Riemann's curvature tensors.
%Thus, 
we can write
\begin{align}\label{tfsu}
\tilde{\t}_{\m\n}=a^\a_{\ \m}\t_{\a\b}a^\b_{\ \n}.
\end{align}
Now, by substitution $y^\m$ with $\yh^\m$ and Eq.(\ref{amiu3}) in Eq.(\ref{tfsu}), we get coordinate dependency of $\tilde{\t}(\yh)$
%We can also achieve coordinates dependency of $\tilde{\t}(\yh)$ by substituting Eq.(\ref{amiu3}) into Eq.(\ref{tfsu}).
%That is:
\begin{align}\label{tfsu2}
\tilde{\t}_{\m\n}=
\t_{\m\n}-\frac{\xi^\a_\n}{6}\t_{\m\a}R^\a_{\ l\n k}\yh^l \yh^k-\frac{\xi^\a_\m}{6}\t_{\a\n}R^\a_{\ l\m k}\yh^l \yh^k.
\end{align}
When $\t^{0i}=0$, the special case of Eq.(\ref{tfsu2}) becomes
\begin{align}\label{tfsu6}
\tilde{\t}_{ij}=
\t_{ij}-\frac{1}{6}\t_{ik}R^k_{\ lj m}\yh^l \yh^m-\frac{1}{6}\t_{kj}R^k_{\ li m}\yh^l \yh^m.
\end{align}

\textbf{Higgs mapping-}
In this section, we set a local framework on a time-independent small subset of $\mathit{M}$ at an event point $"P"$ of space-time. 
Then, global coordinate $"x"$ belonging to $\mathit{M}$ can be written in term of the coordinate of $"P"$ also local coordinate $"y"$.
% limited to the $"p"$ relevant neighborhood.
The coordinates $"y^\m"$ can span the small neighborhood around of $'P'$.
Therefore, 
\begin{align}\label{gtl}
x^\m=
\cr
Y^\m_p+y^\m+A^\m_{\a\b}(p)y^\a y^\b+B^\m_{\a\b,\gamma}(p)y^\a y^\b y^\gamma+\cdots.
\end{align}
Where $'Y_P'$ is a general coordinate of $'P'$.
Since, Eq.(\ref{eq2}) derived from the definition of Eq.(\ref{gtl}),
so $A^\m_{\a\b}(p)$ is zero.
By excluding the upper-order terms, Eq.(\ref{gtl}) becomes:
\begin{align}\label{gt2}
x^\m=Y^\m_p+y^\m+B^\m_{\a\b\gamma}(p)y^\a y^\b y^\gamma.
\end{align}
%$B^\m_{\a\b\gamma}$ is a combination of Riemann's curvature tensors, $B^\m_{\a\b\gamma}=\sigma(R^\m_{\a\b\gamma})$.
%$\sigma(R^\m_{\a\b\gamma})=R^\m_{\ \a\b\gamma}-R^\m_{\ \gamma\a\b}$, in which $\b$ can takes the value $\m$ only. 
%can be evaluated by Higgs approach for the Schwarzschild universe. 
$B^\m_{\a\b\gamma}$ is a combination of Christopher coefficients, $B^\m_{\a\b\gamma}=\l^\m\Gamma^\m_{\a\b,\g}$,
%$\sigma(R^\m_{\a\b\gamma})=R^\m_{\ \a\b\gamma}-R^\m_{\ \gamma\a\b}$, 
in which $\l^\m$ 
%can takes the value $\m$ only. 
calculated for Schwarzschild universe by Higgs approach \cite{higgs}. 
By this choice, Christopher coefficients independent from time.
%the independent from the time.
Therefore, for Schwarzschild universe,
\begin{align}
x^\m=Y^\m(P)+y^\m+\l^\m\Gamma^\m_{\a\b,\n}(P)y^\a y^\b y^\n.
\end{align}
The coefficients $\l^0$ and $\l^i$'s get the values by comparing 
%the A and B. can now be determined by comparing the obtained result from; 
$dx^\n\eta_{\n\m}dx^\m=\{-1-4\l^0\Gamma^0_{m0,n}y^my^n\}dy^0dy^0+\{1+2\l^1\Gamma^1_{00,1}\cdots\}dy^1dy^1+\cdots$ and Eq.(\ref{eq2}).
Our calculations show that, $\l^{\mathrm{temporal}}=-\frac{1}{4}$ and $\l^{\mathrm{spatial}}=\frac{1}{2}$.
With the condition: $\t^{0i}=0$, we have
\begin{align}\label{tfsu3}
\tilde{\t}^{ij}=\t^{ij}-\frac{1}{6}\t^{kj}R^i_{\ mkn}y^my^n-\frac{1}{6}\t^{ik}R^j_{\ mkn}y^my^n,
\end{align}
%Obtained generalized non-commutativity is 
which specialized for Schwarzschild universe.
One can see that $\tilde{\t}^{ij}$ in Eq.(\ref{tfsu3}) is an anti-symmetric tensor,
because they transform by the tetrad approach.

\textbf{Dynamic case of $\mathit{T}_p\mathit{M}$}- 
Due to general relativity, the curvature of space-time affected by massive bodies.
Also, the gravitational waves are one of the solutions to Einstein's equation, 
which occurs due to the change in the distance of the two massive bodies.
So, gravitational waves are ripples in the curvature of space-time which generated by gravitational interactions.
These waves are coming from the depths of space and outward from their source at the speed of light. They have various amplitudes and frequencies.
We especially assume a state that the space-time filled with weak gravitational effects. 
These effects are static waves and depend on time only.   
Gravitational waves explained by the assumption of small fluctuation in the metric tensor \cite{lab16,lab17,lab18,lab19}.
They propagate in the $z$-direction and is at least time and no direction dependent.
In this case, we decompose the metric tensor as:
\begin{align}\label{lgw-metric}
ds^2=\eta_{\m\n}dx^\m dx^\n+h_{ij}(t)dx^i dx^j,
\end{align}
Latin indices run from 1 to 3 while Greek take on the values 0,1,2,3.
Indeed, the first order perturbation correction $h_{ij}$ is a function of coordinates which we assume to be only time dependent.  
The simple form of gravitational background is as follows:
\begin{align}
h_{\m\n}=\left(
\begin{array}{cccc}
0 & 0 & 0 & 0 \\
0 & 0 & h_{12} & 0 \\
0 & h_{21} & 0 & 0 \\
0 & 0 & 0 & 0
\end{array}
\right),
\end{align}
with $\mathfrak{D}^\m h_{\m\n}=0$, where $\mathfrak{D}^\m$ denoting the covariant derivative.
By setting $h_{12}=h_{21}$ as $"\times"$ polarization
are familiar \cite{lab19, lab20}.
For the explaining of the frame field of $\mathit{T}_p\mathit{M}$, it is necessary to use a set of localized coordinates.
We mention that $\mathit{T}_x\mathit{M}$ contains a first order of Riemann's curvature tensor. 
%For this purpose, we can present the following vierbein time dependent fields:
The non zero components of the Christopher connections are $\Gamma^0_{12}$, $\Gamma^1_{02}$ and $\Gamma^2_{01}$.
Also, only non zero independent component of Riemann's curvature tensor is; $R^
1_{020}=+\frac{1}{2}\ddot{h}_{12}$. 
It is possible to find vierbein field based on the modified version of the relation found in Ref.\cite{lab14, lab15}.
In Ref.\cite{lab14,lab15} has offered a relation to make the vierbein field. 
This validity is only for the static state of $\mathit{T}_p\mathit{M}$. 
So that, for the case of gravitational waves, we have to change it.
Our finding vierbein field satisfies the corresponding motion equation: $\mathfrak{D}^\m a^\a_\n=\p_\m a^\a_\n-a^\a_\sigma \Gamma^\sigma_{\m\n}=0$.
Also, we know that the non-zero elements of Christopher connections are $\Gamma^1_{02}=\Gamma^2_{01}=-\frac{\dot{h}}{2}$.
So,  solving the equation of  motion of time-dependent only vierbein fields are $a^\a_\n=\int\ dt\ a^\a_\sigma\Gamma^\sigma_{0\n}$ which reduced to $a^2_1=a^2_2\Gamma^2_{01}=-\frac{\dot{h}}{2}a^2_2$
and $a^1_2=a^1_1\Gamma^1_{02}=-\frac{\dot{h}}{2}a^1_1$. 
In the absence of gravity, we find coordinates such that $a^\a_{\ \m}(y)=\delta^\a_{\ \m}$.
So that, we can set $a^\a_\m=\delta^\a_\m+\Gamma^\a_{0\m}$. 
In the language of Riemann's curvature tensors, we present components of the vierbein field: $$a^\a_{\ \m}=\delta^\a_\m-\int^{\acute{t}}\int^{\acute{\acute{t}}}d\acute{t}d\acute{\acute{t}}R^\a_{0\m 0}(\acute{\acute{t}})$$
whose form is as follows
\begin{align}\label{amiu223}
a^\a_{\ \m}=\left(\begin{array}{cccc}
                 1 & 0 & 0 & 0  \\
                 0 & 1 & -\frac{h}{2} & 0  \\
                 0 & -\frac{h}{2} &  1 & 0  \\
                 0 & 0 &  0 & 1
               \end{array}
\right).
\end{align}
%Because we can check $\mathfrak{D}_\m a^\a_{\ \n}=\p_\m a^\a_{\ \n}-a^\a_{\ \sigma}\Gamma^\sigma_{\m\n}$ and all of the cases vanish directly.  
%%Because, 
%for $\m=0$, $\a=1$ and $\n=2$ (only non zero elements), 
%%we can check $\mathfrak{D}_\m a^\a_{\ \n}=\p_\m a^\a_{\ \n}-a^\a_{\ \sigma}\Gamma^\sigma_{\m\n}$ vanishes
%%%%%So, we have
%%%%%\begin{align}
%%%%%\mathfrak{D}_0 a^1_{\ 2}=\p_0\frac{h}{2}-1*\Gamma^2_{01}=0.
%%%%%\end{align}
%%and all of the cases directly.
%Such as that for $\m=0$, $\a=2$ and $\n=1$, we have $\mathfrak{D}_0 a^2_{\ 1}=0$.
%And for the rest of the components, it is also provable.
In this way, we can replicate Eq.(\ref{tfsu}); $\tilde{\t}_{\m\n}=a^\a_\m\t_{\a\b}a^\b_\n$.
With direct substitution and in the matrix representation, we have
\begin{widetext}
\begin{align}
\left(\begin{array}{cccc}
                 0 & \tilde{\t}_{01} & \ \tilde{\t}_{02} & \ \tilde{\t}_{03}  \\
                 \tilde{\t}_{10} & 0 & \ \tilde{\t}_{12} & \ \tilde{\t}_{13}  \\
                 \tilde{\t}_{20} & \ \tilde{\t}_{21} &  0 & \ \tilde{\t}_{23}  \\
                 \tilde{\t}_{30} & \ \tilde{\t}_{31} &  \ \tilde{\t}_{32} & 0
               \end{array}
\right)
=
%\cr
%\left(\begin{array}{cccc}
%                 1 & 0 & 0 & 0  \\
%                 0 & 1 & 0 & 0  \\
%                 0 & h &  1 & 0  \\
%                 0 & 0 &  0 & 1
%               \end{array}
%\right)\left(\begin{array}{cccc}
%                 0 & \t^{01} & \t^{02} & \t^{03}  \\
%                 \t^{10} & 0 & \t^{12} & \t^{13}  \\
%                 \t^{20} & \t^{21} &  0 & \t^{23}  \\
%                 \t^{30} & \t^{31} &  \t^{32} & 0
%               \end{array}
%\right)\left(\begin{array}{cccc}
%                 1 & 0 & 0 & 0  \\
%                 0 & 1 & h & 0  \\
%                 0 & 0 &  1 & 0  \\
%                 0 & 0 &  0 & 1
%               \end{array}
%\right)
%\cr
\left(\begin{array}{cccc}
                 0 & \ \t_{01}-\frac{h}{2}\t_{02} & \ -\frac{h}{2}\t_{01}+\t_{02} & \ \t_{03}  \\
                 \t_{10}-\frac{h}{2}\t_{20} & \ 0 & \ \t_{12} & \ \t_{13}-\frac{h}{2}\t_{23}  \\
                 -\frac{h}{2}\t_{10}+\t_{20} & \ \t_{21} &  \ 0 & \ -\frac{h}{2}\t_{13}+\t_{23}  \\
                 \t_{30} & \ \t_{31}-\frac{h}{2}\t_{32} &  \ -\frac{h}{2}\t_{31}+\t_{32} & \ 0
               \end{array}
\right)
\end{align}
\end{widetext}
Therefore, in the presence of gravitational waves, the non-commutativity dependency on coordinates is achievable.
Here, we should use the extension of the metric interrupt. 
%$h(t)=h(Y_p)+\dot{h}(Y_p)t+\frac{1}{2}\ddot{h}(Y_p)tt+\cdots$.
In this manner, the obtained result simplified to
%\begin{widetext}
%\begin{align}
%\left(\begin{array}{cccc}
%                 0 & \tilde{\t}^{01} & \ \tilde{\t}^{02} & \ \tilde{\t}^{03}  \\
%                 \tilde{\t}^{10} & 0 & \ \tilde{\t}^{12} & \ \tilde{\t}^{13}  \\
%                 \tilde{\t}^{20} & \ \tilde{\t}^{21} &  0 & \ \tilde{\t}^{23}  \\
%                 \tilde{\t}^{30} & \ \tilde{\t}^{31} &  \ \tilde{\t}^{32} & 0
%               \end{array}
%\right)
%=
%\left(\begin{array}{cccc}
%                 0 & \ \t^{01}-\t^{02}R^0_{120}t^2 & \ -\t^{01}R^0_{120}t^2+\t^{02} & \ \t^{03}  \\
%                 \t^{10}-\t^{20}R^0_{120}t^2 & \ 0 & \ \t^{12} & \ \t^{13}-\t^{23}R^0_{120}t^2  \\
%                 -\t^{10}R^0_{120}t^2+\t^{20} & \ \t^{21} &  \ 0 & \ -\t^{13}R^0_{120}t^2+\t^{23}  \\
%                 \t^{30} & \ \t^{31}-\t^{32}R^0_{120}t^2 &  \ -\t^{31}R^0_{120}t^2+\t^{32} & \ 0
%               \end{array}
%\right)
%\end{align}
%\end{widetext}
\begin{align}
\tilde{\t}_{\m\n}=
\t_{\m\n}
\cr
-\t_{\m\b}\int^{\acute{t}}\int^{\acute{\acute{t}}}d\acute{t}d\acute{\acute{t}}R^\b_{0\n 0}(\acute{\acute{t}})-\t_{\b\n}\int^{\acute{t}}\int^{\acute{\acute{t}}}d\acute{t}d\acute{\acute{t}}R^\b_{0\m 0}(\acute{\acute{t}}).
\end{align}
in which, $\int^{\acute{t}}\int^{\acute{\acute{t}}}d\acute{t}d\acute{\acute{t}}R^\b_{0\m 0}(\acute{\acute{t}})=\frac{h_{\b\m}(t)}{2}$.

\section{Conclusion}

This work is in the first order of Riemann's curvature tensors and at the level of tangential space.  
In these limits, by employing vierbein field, we found coordinates dependency of generalized non-commutative relationship.
We made local fields which satisfied the complement relation of Eq.(\ref{vierbein-fields}).
So, we derived the coordinates dependency for other than the Minkowski space-time.     
We specialized the generalized non-commutativity coordinates relationship for Schwarzschild universe as well as for space-time in the presence of weak gravitational waves.
Also, we showed that behavior of the coordinates non-commutativity like as a tensor.
Based on Higgs approach, we found a different way to find the coordinates non-commutativity dependence on local coordinates.
Obtained results showed that these methods are comparable.

\section{Acknowledgments}
The author thanks Shahrekord University for supporting this work with a research grant.
\newline
%%%%%%%%%%%%%%%%%%%%%%%%%%%%%%%%%%%%%%%%%%%%%%%%%%


\begin{thebibliography}{99}

%Relativity and General Relativity and Vierbein Fields

%29
\bibitem{lab01} L. Parker, D. J. Tomas, Quantum Field Theory in Curved Space-time, (London: Cambridge University Press, 2009), p. 144-151
%10
\bibitem{lab02} Hans C. Ohanian, Gravitation and Space-time, (New York: W. W. Norton and Company, 1976), p. 141-150
%30
\bibitem{lab03} N. D. Birrell, P. C. W. Davies, Quantum Fields in Curved Space, (London: Cambridge University Press, 1982), p. 81-88
%31
\bibitem{lab04} S. Weinberg, Gravitation and cosmology, (New York: John Wiley and Sons, Inc., 1972),  
%32
\bibitem{lab05} F. De Felic, C. J. S. Clarke, Relativity on curved manifold, (New York: Cambridge University Press, 1990),


%Non Commutativity
%16
\bibitem{lab06} Lecture. Notes in Phys, Non-commutative Space-time: Symmetries in Non-commutative Geometry and Field Theory, Vol. 774, P. Aschieri, M. Dimitrijevic, P. Kulish, et al, (Berlin Heidelberg: Springer, 2009), P. 1-4
%%%P. Aschieri, M. Dimitrijevic, P. Kulish, et al, Non Commutative Spacetimes: Symmetries in Non Commutative Geometry and Field Theory,
%%%Lecture. Notes in Phys, (Berlin Heidelberg: Springer, {\bf 774}, 2009), P. 1-4
%11
\bibitem{lab07} A. Connes, M. Marcolli, Non-commutative Geometry, Quantum Fields and Motives, (London: Academic Press, 1994), p. 1-31
%15
\bibitem{lab08} D. J. Gross, N. A. Nekrasov, JHEP, {\bf 0103}: 044 (2001)
%12
\bibitem{lab09} R. J. Szabo, Phys. Rept, {\bf 378}: 207-299 (2003) 
%13
\bibitem{lab10} A. Fischer, R. J. Szabo, JHEP, {\bf 0902}: 031 (2009)
%14
\bibitem{lab11} N. Seiberg, E. Witten, JHEP, {\bf 9909}: 032 (1999)
%25
\bibitem{lab12} A. Jafari, Eur. Phys. J. C, {\bf 73}: 2271 (2013)
%Schwarzschild
%9
\bibitem{lab13} R. D'inverno, Introducing Einstein's Relativity, (New York: Oxford University Press Inc., 1993), p. 269-285
%5
\bibitem{lab14} L. Parker, Phys. Rev. Let, {\bf 44}: 1559 (1980)
%6
\bibitem{lab15} L. Parker, Phys. Rev. D, {\bf 22}: 1922 (1980)
%7
\bibitem{higgs} P. W. Higgs, J. Phys. A, Math. Gen, {\bf 12}: 309 (1994)
%WGW
\bibitem{lab16} J. Weber, General Relativity and Gravitational Waves, Dover Publications (Mineola park, New York: Interscience Publisher Inc., 1961), p. 87-144
%8
\bibitem{lab17} C. W. Misner, K. S. Thorne, J. A. Wheeler, Gravitation, (San Francisco: Freeman Publishing Company, 1973), p. 435-445
%4
\bibitem{lab18} M. Maggiore, Gravitational Waves, (New York: Oxford University Press, Inc., 2008),p. 42-56
%33
\bibitem{lab19} A.D. Speliotopoulos, Phys. Rev. D, {\bf 51}: 1701-1709 (1995)
%34
\bibitem{lab20} A. Saha, S. Gangopadhyay, S. Saha, Phys. Rev. D, {\bf 83}: 025004 (2011)
%

\end{thebibliography}
\end{document}